\documentclass[10pt,conference]{IEEEtran}
\pdfoutput=1

\usepackage[colorlinks,allcolors=black,linktocpage=true]{hyperref}
\usepackage{algorithmic}
\usepackage{cite}
\usepackage{cleveref}
\usepackage{graphicx}
\usepackage{lipsum}
\usepackage{makecell}
\usepackage{siunitx} 
\usepackage{textcomp}
\usepackage{todonotes}
\usepackage{xcolor}
\usepackage{xspace}

\begin{document}

\title{Is This You, LLM? Recognizing AI-written Programs with Multilingual Code Stylometry}

\author{\IEEEauthorblockN{Andrea Gurioli}
  \IEEEauthorblockA{\textit{DISI} \\
    \textit{University of Bologna}\\
    Bologna, Italy \\
    andrea.gurioli5@unibo.it}
  \and
  \IEEEauthorblockN{Maurizio Gabbrielli}
  \IEEEauthorblockA{\textit{DISI} \\
    \textit{University of Bologna}\\
    Bologna, Italy \\
    maurizio.gabbrielli@unibo.it}
  \and
  \IEEEauthorblockN{Stefano Zacchiroli}
  \IEEEauthorblockA{\textit{LTCI, Télécom Paris} \\
    \textit{Institut Polytechnique de Paris}\\
    Palaiseau, France \\
    stefano.zacchiroli@telecom-paris.fr}
}

\maketitle

\def\DataHumanThreshold{630\xspace}
\def\DataMultiProvThreshold{470\xspace}
\def\DataTopLanguages{10\xspace}
\def\DataMultiProvenance{90\xspace}
\def\DataSnippetsMp{\num{121 247}\xspace}
\def\DataAvgAccuracy{84.1\%\xspace}
\def\DataStddev{3.8\%\xspace}
\def\Dataset{H-AIRosettaMP\xspace}
\def\DataRosettaSnippets{\num{79 013}\xspace}
\def\DataHAIRosettaMPTasks{1127\xspace}
\def\DataRosettaLangs{883\xspace}
\def\DataRosettaTasks{1203\xspace}
\def\DataLanguages{C++, C, C\#, Go, Java, JavaScript, Kotlin, Python, Ruby, Rust\xspace}

\begin{abstract}
With the increasing popularity of LLM-based code completers, like GitHub Copilot, the interest in automatically detecting AI-generated code is also increasing---in particular in contexts where the use of LLMs to program is forbidden by policy due to security, intellectual property, or ethical concerns.

  We introduce a novel technique for \emph{AI code stylometry}, i.e., the ability to distinguish code generated by LLMs from code written by humans, based on a transformer-based encoder classifier.
  Differently from previous work, our classifier is capable of detecting AI-written code across 10 different programming languages with a single machine learning model, maintaining high average accuracy across all languages (\DataAvgAccuracy$\pm$\,\DataStddev).

  Together with the classifier we also release \Dataset, a novel open dataset for AI code stylometry tasks, consisting of \DataSnippetsMp code snippets in 10 popular programming languages, labeled as either human-written or AI-generated.
  The experimental pipeline (dataset, training code, resulting models) is the first fully reproducible one for the AI code stylometry task.
  Most notably our experiments rely only on open LLMs, rather than on proprietary/closed ones like ChatGPT.
\end{abstract}

\begin{IEEEkeywords}
  code stylometry, large language models, AI detection, code generation, data provenance, deep learning \end{IEEEkeywords}

\section{Introduction}
\label{sec:introduction}

LLM-based code completers~\cite{codex, codellama, starcoder} (or \emph{code LLMs} for short in this paper), as exemplified by GitHub Copilot\footnote{\url{https://github.com/features/copilot/}, accessed 2024-09-24}, are becoming popular automatic programming tools among software developers. Preliminary evaluations of code LLM results show that they can produce either correct or buggy code~\cite{dakhel2023copilot_eval, sandoval2023code_llm_security}, depending on how they are used.
Specifically, code LLMs can be useful assets for expert programmers who quickly learn to use them well or a liability for novice developers who lack the experience to skip misleading answers quickly.
The real impact of code LLMs on developer productivity also remains unclear, with growing interest in defining proper metrics to evaluate it~\cite{agarwal2024copilot_eval}.

Similarly, policy-wise, the use of code LLMs can be frowned upon or outright forbidden, depending on the context.
Security- and privacy-sensitive environments might forbid the use of code LLMs hosted by 3rd parties---like Copilot, hosted by GitHub, or ChatGPT\footnote{\url{https://openai.com/chatgpt/}, accessed 2024-09-24} by OpenAI---to avoid leaking internal code in prompts.
(Self-hosted open-weight LLMs, like Code Llama~\cite{codellama} and StarCoder~\cite{starcoder} mitigate this issue.)
In teaching contexts, such as schools and universities, the use of code LLMs can be considered cheating (depending on the assignment goals), with severe consequences for the students who use them~\cite{hoqCS_course}.

Legal and licensing risks are also ongoing concerns when using code LLMs~\cite{ren2024copyright}.
Even leaving aside the hot legal topic of whether training LLMs on third-party unlicensed material is allowed (or ethical), code LLMs can output verbatim parts of their training datasets, a phenomenon known as \emph{recitation}~\cite{ziegler2021copilot_recitation}, which might expose their \emph{users} to legal liabilities~\cite{bukariFirst} if generated code is integrated into a product put on the market.

\subsection{Problem statement}

These practical needs have spawned an interest in \emph{automatically recognizing code generated by LLMs}, distinguishing it from code written by humans.
This is an instance of the more general task of automatically detecting \emph{who} wrote a given piece of code, known in the literature as \emph{code stylometry} (or \emph{code authorship attribution}, or \emph{code author recognition})~\cite{oman1989FirstStyl, caliskan2015stylAST, bogomolov}.

Previous work~\cite{bukariFirst, li_human_ai_styl, hoqCS_course, chatGPT_code_det_oedingen, Human_AI_python} has already applied code stylometry techniques to the recognition of ``AI authors'' (i.e., code LLMs), with three recurrent characteristics: (1) detection is possible on a single programming language \emph{at a time}; (2) the tested code LLM is a proprietary, non-open tool or model (e.g., Copilot, or ChatGPT), which hinders scientific reproducibility and replicability; (3) detection is based on traditional machine learning techniques (e.g., random forest classifications, LSTM, or code2vec embeddings).

The goal of this paper is to improve the state-of-the-art of the detection of AI-written programs, by addressing the following research question:
\begin{itemize}
  \itshape
\item[] \textbf{RQ1}: Is it possible to detect source code generated by code LLMs, achieving high accuracy across several different programming languages?
\end{itemize}
Answering this question affirmatively would improve over limitation (1) above, which can be particularly annoying in projects where multiple programming languages are in use, as it is often the case.
Methodologically, we aim to answer RQ1 following a fully reproducible experimental approach (addressing limitation 2 above) and using more recent machine learning techniques (point 3 above) that, as we will see, are needed to achieve good results in the desired multilingual setting.

\subsection{Contributions}

With this paper we make the following novel contributions:
\begin{enumerate}

\item We release a novel, balanced, open \textbf{dataset for the AI code stylometry task}, which contains \DataSnippetsMp code snippets in total, written in 10 different popular programming languages: \DataLanguages.
  Each snippet is labeled as either having been authored by a human (to solve a specific task in the context of the Rosetta Code project~\cite{rosetta-code}) or as generated by StarCoder2~\cite{starcoder2} (a state-of-the-art open code LLM), via code translation from a (human-authored) snippet written in a different programming language to solve the same task.

\item We train a \textbf{transformer-based encoder classifier}---a novel architecture for the AI code stylometry task---on the above dataset.
  Using it we answer the stated research question affirmatively, showing that it is possible to recognize multilingual AI-generated code, across 10 popular programming languages, with an average accuracy of \DataAvgAccuracy ($\pm\,\DataStddev$).

\item We release an \textbf{open source tool} based on the trained classifier, which can be used to detect whether code snippets of interest have been AI-generated or not.
  The tools is available both as a hosted version on Hugging Face\footnote{\url{https://huggingface.co/spaces/isThisYouLLM/Human-Ai}} and as a command line (CLI) tool distributed with this paper replication package.

\item All our experiments are \textbf{fully reproducible}: the initial dataset is openly available and can be regenerated using Rosetta Code data and StarCoder2; the training and evaluation pipeline is available as part of the replication package of this paper (see the Data availability statement at the end).

\end{enumerate}

\section{Methodology}
\label{sec:methodology}

\begin{figure*}
  \centering
  \includegraphics[width=\linewidth]{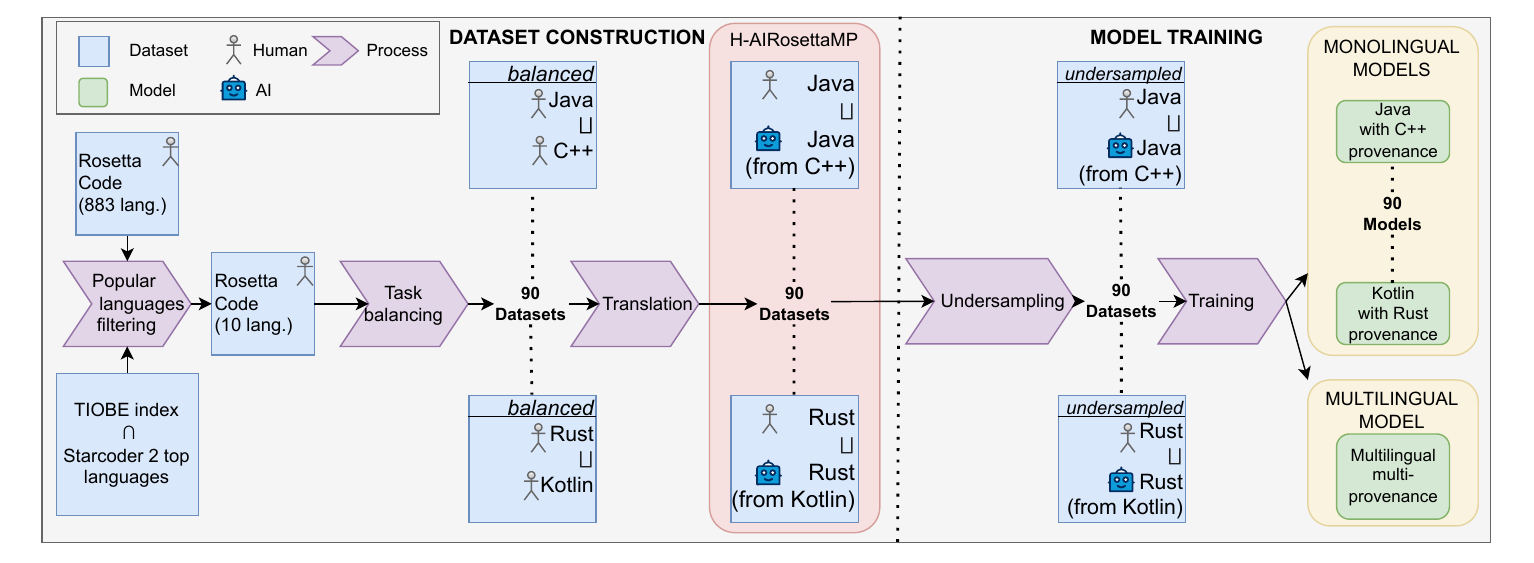}
\caption{Experimental methodology. The process is divided into two main steps: (1) The \textit{Dataset construction}, which starts from the filtered Rosetta 
Code dataset and terminates in the \emph{\Dataset}, obtained via code translation, comprising 90 (sub-)datasets. Each dataset is labeled by the author (Human or AI) and is represented by dst (the language of the dataset) and src language (the provenance of the AI-generated part of the dataset); (2) The \textit{model training}, that shows the process leading to 90 monolingual models (one per dataset) and 1 multilingual model.}
  \label{fig:methodology}
\end{figure*}

The experimental methodology followed for this work is depicted in \Cref{fig:methodology}.
It consists of two parts: (1) dataset construction, described next, leading to the creation of the \Dataset dataset; (2) model training, described in \Cref{sec:human-ai}, leading to multiple classifier models, whose performances are analyzed in \Cref{sec:results}.

\subsection{Dataset construction}
\label{sec:dataset-construction}

Our goal is to train a machine-learning classifier that can distinguish human-written from AI-generated code, across several programming languages (RQ1).
To that end we need first and foremost a training dataset, covering multiple languages, and containing snippets labeled as either human-written or AI-generated.
To the best of our knowledge such a dataset did not exist before, so we set to create one.
While doing so, we pursued the methodological goals of making its construction fully reproducible and releasing it open data.

\subsubsection{Programming language selection}

As a starting point for human-written code snippets, we used Rosetta
Code~\cite{rosetta-code}, a programming chrestomathy project that collects and
publishes solutions to the same programming tasks in as many different languages
as possible, to showcase similarities and differences across languages.
We retrieved a version of the Rosetta Code timestamped as July 1st, 2022.
The retrieved dataset contained \DataRosettaSnippets code snippets, each representing a solution to one among \DataRosettaTasks programming tasks in total, written in one among \DataRosettaLangs programming languages. 

In order to both respond to real-world use cases and maximize data availability for the later training phase, we selected our target programming languages for AI code stylometry based on their popularity.
To rank languages by popularity, we retrieved the TIOBE index~\cite{tiobe} ranking, as of May 2024.
From the TIOBE ranking we removed all languages not present in the training dataset of the open code LLM used in our experiments, namely StarCoder2-15B~\cite{starcoder2} (see later in this section for a discussion of our choice of LLM).
To conclude this step (\emph{Popular languages filtering} in \Cref{fig:methodology}), we selected the top-10 remaining languages by ranking order---10 languages being a very significant step forward w.r.t.~the state of the art of AI code stylometry performed on at most 2 languages at a time.

We hence obtained a set of 10 popular programming languages, together with hand-written snippets in those languages from Rosetta code, that are also well-known to the code LLM used later to generated AI-authored snippets: \DataLanguages.

\smallskip
Before dwelling into details, here are the two key intuitions behind the next steps:
\begin{enumerate}

\item \emph{Human-written snippets} in the target dataset are unmodified snippets coming from Rosetta Code: they were all contributed as task solutions by humans participating in the initiative.

\item \emph{AI-written snippets} in the target dataset are generated by a code LLM (specifically: StarCoder2~\cite{starcoder2}) using \emph{cross-language translation} form a source programming language \emph{src} (called the \emph{provenance language} in the following) to a destination language \emph{dst}, as previously done by Li et al.~\cite{li_human_ai_styl}.
  Input to the translation is a human-written snippet coming from Rosetta Code (as per (1) above); output of the translation is an AI-written snippet (by StarCoder2) that will be integrated into the target dataset.
  Details on the translation step are provided later in this section.

\end{enumerate}

\subsubsection{Task balancing}

To avoid skewing the AI-written part of the dataset by translating from a single source programming language (which might be more affine to one target language than another), all \emph{human-written} snippets in the dataset for a given programming language \emph{src} have been translated to \emph{all other 9 languages} among the 10 selected languages.
This constitute a total of \DataMultiProvenance ($= 10 \times 9$) sub-datasets, each formed by a $\{\mathit{src}, \mathit{dst}\}$ unordered language pair, where $\mathit{src}\neq\mathit{dst}$.

The initial \DataMultiProvenance (sub-)datasets (before the \emph{Translation} step in \Cref{fig:methodology}) have been obtained by selecting snippets from Rosetta Code in a way that created balanced datasets.
Specifically, in each $\{\mathit{src}, \mathit{dst}\}$ dataset, we only kept Rosetta Code snippets pertaining to the same task.
That is, each solution written in programming language \emph{src} is kept in the dataset $\{\mathit{src}, \mathit{dst}\}$ if and only if a solution \emph{for the same task} exists also for programming language \emph{dst}, and vice-versa.

After this step, we obtained the balanced \DataMultiProvenance (sub-)datasets shown in \Cref{fig:methodology} just before the \emph{Translation} step that we describe next.

\subsubsection{Translation}

Li et al.~\cite{li_human_ai_styl} pioneered using code translation for building the AI-generated part of datasets for AI code stylometry.
They discussed three alternative methodologies to do so:
\begin{enumerate}

\item \emph{Code translation} involves providing the generative model with a code snippet in one programming language, asking the model to translate it into a different language;

\item \emph{Functional translation} involves providing a natural language description of the desired task, asking to generate a solution;

\item \emph{Functional customization} involves providing an existing snippet of code, asking to provide an explanation of what it does first, and then asking to generate a solution based on the description.

\end{enumerate}
We considered all three options for our needs and concluded that (2) and (3) are not suitable option, because in our evaluation (using various code LLMs) they often end up producing ``skeleton code'', with holes that remain to be filled by the user.
Keeping those incomplete snippets in the target dataset would give an unfair advantage to the AI detector, because they will be fairly easy to distinguish from complete code snippets from Rosetta Code (written by humans).
We then settled for code translation (1) and applied it to the \DataMultiProvenance balanced datasets obtained from the previous step.

\begin{figure*}
  \centering
  \includegraphics[width=\linewidth]{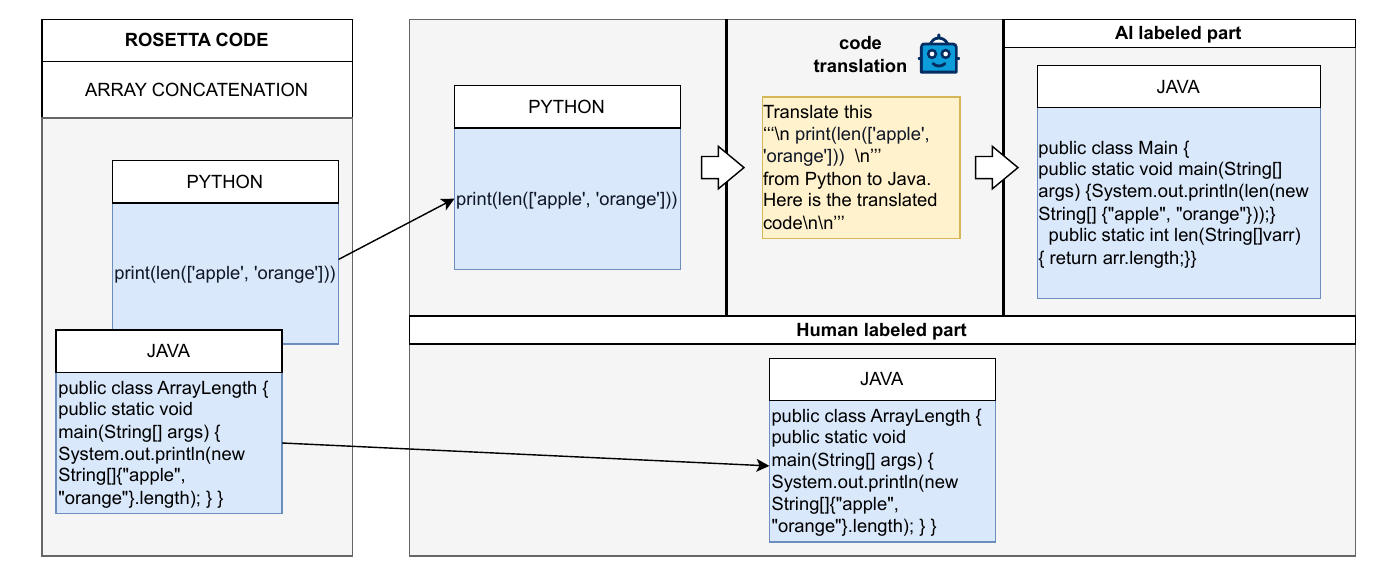}
  \caption{Code translation step. The human-labeled part of the dataset (the \texttt{ArrayLength} Java class here) is a solution to a task from Rosetta Code (Array concatenation). The AI-labeled part is obtained via code translation (from Python to Java) using StarCoder 2. Input to the translation is a human-written solution for the same task, in a different programming language.}
  \label{fig:translation}
\end{figure*}

Specifically, for each (sub-)dataset $\{\mathit{src}, \mathit{dst}\}$, we took all the snippets in it written in the \emph{dst} programming language, and used the StarCoder2~\cite{starcoder2} generative model to translate the snippet into the \emph{src} programming language (\emph{Translation} step in \Cref{fig:methodology}, further detailed in \Cref{fig:translation}).

The choice of StarCoder2 as code completion model is due to its being an open model, both in its weights (available for download and reuse under the terms of the Open RAIL-M v1 license) and in its training dataset (obtained from the Software Heritage archive~\cite{softwareHeritage}).
Openness is a strong requirement to achieve our goal of full reproducibility of the experimental pipeline, which would not be achievable using closed models such as Copilot or ChatGPT (and indeed has not been achieved in previous work in the literature).
StarCoder2 achieves 46.3\% accuracy in the HumanEval benchmark~\cite{codex}, a widely used benchmark for assessing the coding abilities of generative models, making it outperform coding models like CodeLlama and DeepSeekCoder~\cite{codellama, deepseek}.
In summary: StarCoder2 is the best performing code LLM among those that are open enough to satisfy the reproducibility requirement.

\begin{figure}
  \raggedright
  \texttt{Translate this  ```\textbackslash n CODE\_SNIPPET \textbackslash n''' from SOURCE\_LANGUAGE to TARGET\_LANGUAGE. Here is the translated code\textbackslash n\textbackslash n'''}
  \caption{Synopsis of the prompt given to StarCoder2 for translating a given code snippet (\texttt{CODE\_SNIPPET} in the text) from a source programming language (\texttt{SOURCE\_LANGUAGE}) to a target one (\texttt{TARGET\_LANGUAGE}).
    A prefix of the desired answer (\texttt{Here is the\ldots}) is provided because StarCoder2 has not been fine-tuned for chat-based interaction and is strictly a completion model.}
  \label{fig:prompt}
\end{figure}

To translate a snippet from language \emph{src} to \emph{dst}, we give to StarCoder2 the prompt whose synopsis is shown in \Cref{fig:prompt}.
When reading the generated output we select each next token by taking the one with the highest likelihood (greedy search), as it is commonplace in translation tasks.

Due to memory and computational limitations, we excluded snippet pairs $\{\mathit{src}, \mathit{dst}\}$ that would result in prompts longer than 1024 tokens, and we set 2048 as the maximum length in the generative phase.
We have also excluded snippet pairs for which StarCoder2 returned malformed outputs (e.g., empty or lacking the closing \texttt{'''} delimiter).

After translation, each of the 90 $\{\mathit{src}, \mathit{dst}\}$ sub-datasets is now composed of snippets in a \emph{single} programming language (\emph{dst}), labeled as either human-written (by a Rosetta Code contributor) or AI-written (by StarCoder2 via code translation).

All together, the 90 sub-datasets form a single reproducible dataset, called \emph{\Dataset}, which we release publicly as open data for others to experiment with.
\emph{\Dataset} comprises \DataSnippetsMp{} snippets, with \DataHAIRosettaMPTasks{} unique tasks in \DataTopLanguages{} popular programming languages.

\smallskip
Note how the \Dataset dataset satisfies multiple important requirements for the AI code stylometry task.
As discussed by Caliskan-Islam~\cite{caliskan2015stylAST} as an important feature: it contains multiple snippets, authored by multiple authors (grouped in two, in our case: humans vs AI), implementing different tasks.
Specifically, it is not the case that the two authors are partitioned by task: for each task we have both a human-authored solution and an AI-authored one.
This avoids the risk that the classifier will learn to distinguish \emph{tasks}, rather than authors.
Additionally, the dataset satisfies all the requirements associated with RQ1, namely: it is multilingual (with 10 languages), it is openly available and fully reproducible.

\subsection{Model training}
\label{sec:human-ai}

\subsubsection{Classifier architecture}

Until now, only classical machine learning techniques (discussed in detail in \Cref{sec:related}) have been applied to the AI code stylometry task.
On the other hand, Niu et al.~\cite{empirical_LLMs} showed how transformer-based architectures are state-of-the-art for several \emph{code understanding} tasks.
In the context of natural language (as opposed to code), recent works obtained successful results on AI recognition~\cite{mitrovic2023chatgpt, liaoChatGPT} using analogous architectures.

\begin{figure}
  \centering
  \includegraphics[width=0.7\linewidth]{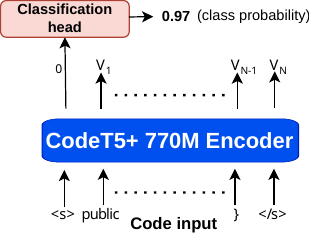}
  \caption{Transformer-based architecture of the Human/AI stylometry classifier.
    Input source code is tokenized and provided as input to the CodeT5plus encoder, which produces as outputs multiple vectorial representations.
    The first token (\texttt{<s>} in the figure) is used as input for the classification head, which produces the final class probability.}
  \label{fig:model}
\end{figure}

In this work, we apply, to the best of our knowledge for the first time, a transformer-based machine learning architecture to the task of AI code stylometry.
Specifically, we use CodeT5plus-770M~\cite{wang2023codet5+}, a pre-trained code transformer architecture, in an encoder setup, as shown in \Cref{fig:model}.
We provide the model with a tokenized text input and obtain several vectorial representations as the number of tokens.
To distinguish human- from AI-written code, we add, following Niu et al.~\cite{empirical_LLMs}, a classification head to the first output representation of the model, corresponding to an always present classification token (see \Cref{fig:model}).
The classification head consists of a linear layer with a ReLU activation function, 20\% dropout, followed by a final linear layer for binary classification.

\subsubsection{Undersampling}

As we describe below, we have trained 91 classifier models in total: one \emph{monolingual model} for each of the 90 $\{\mathit{src}, \mathit{dst}\}$ sub-datasets + one \emph{multilingual model} on a multilingual dataset sampled from the entire \Dataset.

Before training the monolingual models, we undersampled each sub-dataset to the threshold of \DataMultiProvThreshold{} code snippets for each Human/AI class, corresponding to the minimum amount of snippets across all sub-datasets.

Before training the multilingual model, we want to make sure that: (1) AI-written snippets, generated via code translation, come from a uniform distribution of source languages before the translation; (2) for both AI-written and human-written snippets, only a single solution for a given task is present (to avoid learning about the task, rather than learning about the author style).
To ensure these properties, we processed the 10 languages one by one.
For each language \emph{dst}, we collect $\DataMultiProvThreshold+\DataMultiProvThreshold=940$ snippets (half AI-written, half human-written).
When collecting AI-written snippets, we sample across the other 9 provenance languages \emph{src}, with a uniform distribution.
When collecting both AI-written and human-written snippets, we never select more than one solution for the same task; at most, the solution to the same task can hence appear twice in a given programming language, once as human-written and once as AI-written.

As an additional data cleaning step, we also removed all leading and trailing spaces from all code snippets (both human-written and AI-written) because AI-translated snippets exhibit recognizable heading/trailing spacing patterns, and we wanted to avoid unfairly advantaging classifiers that might learn from them (in real-world use cases those spaces would most likely not be preserved as is).

\subsubsection{Training}
 
We adjusted the model hyperparameters, starting from the setup proposed by Wang et al.~\cite{wang2023codet5+}, picking a subset of the dataset (all languages with Python provenance except for the Python language, translated from C++) and validated the model, obtaining a hyper-parameter setup for the rest of the experiments.
We used AdamW~\cite{AdamW} as an optimizer with a weight decay of 0.01. We trained each model for 15 epochs, multiplying the learning rate after 10 epochs by a 0.1 factor, with an initial learning rate of 2e-05.

After training the 90 monolingual models, we observed different results for the same \emph{dst} language, coming from datasets with different provenance language \emph{src} (see \Cref{sec:results} for details).
We inspected this phenomenon by testing the best model for a language \emph{dst} on datasets with different \emph{src} provenance. 

To obtain a model capable of handling several different languages as input, we trained the multilingual model using the entire \Dataset dataset (after undersampling).

All models were trained with an 80\%/20\% training/test split.

We then compared our results with the best-performing results in the literature~\cite{li_human_ai_styl,chatGPT_code_det_oedingen}.
Our classifiers were trained on the novel \Dataset dataset, which is different from datasets used in previous works.
Thus, for a fair comparison, we re-trained Li et al. and Oedingen et al.~\cite{li_human_ai_styl,chatGPT_code_det_oedingen} classifiers over our dataset.

We trained at first four baseline models following Li et al.~\cite{li_human_ai_styl} two methodologies (Random Forest and J48) over our Java and C++ sub-datasets with Kotlin provenance (because it obtained the best performances across all \emph{src} provenances).
The models were trained and evaluated using the same methodology of Li et al., with 10-fold cross-validation.

As our last comparison, we trained a baseline model following the best-performing methodology of Oedingen et al~\cite{chatGPT_code_det_oedingen} over our Python dataset (also with Kotlin provenance).

\subsubsection{Evaluation}

We evaluated each of the trained monolingual classifiers on the respective dataset (in-distribution test), noting down the resulting accuracy.
We evaluated in the same way the trained multilingual classifier on its own dataset (in-distribution test), which contains snippets in all languages and translated from all \emph{src} languages (for the AI-labeled part).

We evaluated the 5 baseline models on different datasets (see \Cref{tab:comp} for reference): RF Java and J48 Java baselines on the Java sub-dataset with translation from Kotlin (best average accuracy among all \emph{src} provenance languages for Java); RF C++ and J48 C++ on the C++ sub-dataset (provenance: Kotlin); XGB-TF-IDF Python on the Python sub-dataset (provenance: Kotlin).
Finally, we evaluated the monolingual models needed for comparison with baselines on out-distribution languages with Kotlin provenance, except for Kotlin itself, where Go provenance was used.

\section{Dataset}
\label{sec:dataset}

The \Dataset dataset is released publicly as part of our replication package (on Zenodo, see Data availability statement at the end of the paper) and also mirrored on Hugging Face.\footnote{\url{https://huggingface.co/datasets/isThisYouLLM/H-AIRosettaMP}}
The dataset comes in tabular form, with one row per snippet, for a total of \DataSnippetsMp rows.
Each column in the table provides some information about the snippet:
\begin{itemize}

\item \emph{\ttfamily task\_name, task\_url, task\_description}: information about the Rosetta Code~\cite{rosetta-code} task that the snippet implements, respectively: the task name (e.g., Array concatenation), URL on the Rosetta Code website (e.g., \url{http://rosettacode.org/wiki/Array_concatenation}), and natural language description of the task.

\item \emph{\ttfamily language\_name}: the programming language in which the code snippet is written, one of: \DataLanguages.

\item \emph{\ttfamily code}: the actual, full source code of the snippet as a string.

\item \emph{\ttfamily target}: a binary label denoting whether the snippet is human- or AI-written.

\item \emph{\ttfamily set}: the name of the specific sub-dataset, e.g., \texttt{"Java\_from\_C++"} for the Java snippet dataset whose AI-written parts were obtained via translation from C++ (the human-written snippets, on the other hand, were natively written in Java).

\end{itemize}

\begin{table*}
  \caption{Average length of code snippets in the \Dataset dataset, by programming language, measured in characters. t-test ($\alpha=0.05$) is computed between the AI-written and Human-written groups}
  \centering
  \label{tab:dataset}
  \begin{tabular}{lrrrrrr} 
    \hline
    \textbf{Language}  & \multicolumn{3}{c}{\textbf{Snippet length} (character avg. $\pm$ std)}& \multicolumn{3}{c}{\textbf{t-test}}\\  
    & \multicolumn{1}{c}{\bfseries All} & \multicolumn{1}{c}{\bfseries AI-written} & \multicolumn{1}{c}{\bfseries Human-written} &\multicolumn{1}{c}{t-statistic}&\multicolumn{1}{c}{95\% CI}  &\multicolumn{1}{c}{p-value}    \\
    \hline
    \textbf{C++} & $1183\pm67$ & $1061\pm73$ & $1306\pm82$ &$6.72$&168 to 323&$<0.01 $ \\ 
    \textbf{C} & $1094\pm57$ & $1077\pm91$ & $1112\pm44$ &$1.04$&-37 to 107& $0.31$ \\
    \textbf{C\#} &  $1248\pm69$ & $1240\pm103$ & $1257\pm76$ &$0.39$& -74 to 107&$0.69$\\ 

    \textbf{Go} & $977\pm63$ & $883\pm79$ & $1072\pm72$ &$5.30$& 114 to 265&$<0.01$ \\ 
    \textbf{Java} & $1262\pm88$ & $1245\pm102$ & $1278\pm101$ &$-0.67$&-134 to 69& $0.51$ \\
    \textbf{JavaScript} & $933\pm60$ & $787\pm82$ & $1079\pm87$ &$7.32$&207 to 377& $<0.01$ \\ 
    \textbf{Kotlin} & $920\pm56$ & $872\pm91$ & $969\pm67$ &$2.60$&18 to 178&$0.02$\\ 
    \textbf{Python} & $744\pm57$ & $729\pm86$ & $761\pm51$ &$0.96$&-38 to 102& $0.35$ \\ 
    \textbf{Ruby} & $584\pm42$ & $650\pm62$ & $518\pm36$ &$-5.52$&-183 to -82& $<0.01$ \\ 
    \textbf{Rust} & $992\pm54$ & $909\pm81$ & $1076\pm71$&$4.63$ &90 to 243& $<0.01$ \\
    \hline
  \end{tabular}
\end{table*}

\noindent
As a simple descriptive statistics, \Cref{tab:dataset} shows the average length of code snippets in the entire dataset by programming language, measured as the number of characters.
We conducted a t-test for each language between the Human and AI-labeled groups of snippets with $\alpha=0.05$ after having tested data normality and variance.
When looking at the breakdown between AI- and human-written snippets, we see that six languages out of ten significantly differ in number of characters ($p<0.05$ \Cref{tab:dataset}).
The noticeable differences in snippet lengths between AI- and human-written code suggest that length could be a predictive feature in code detection models.

\begin{figure}
  \centering
  \includegraphics[width=\linewidth]{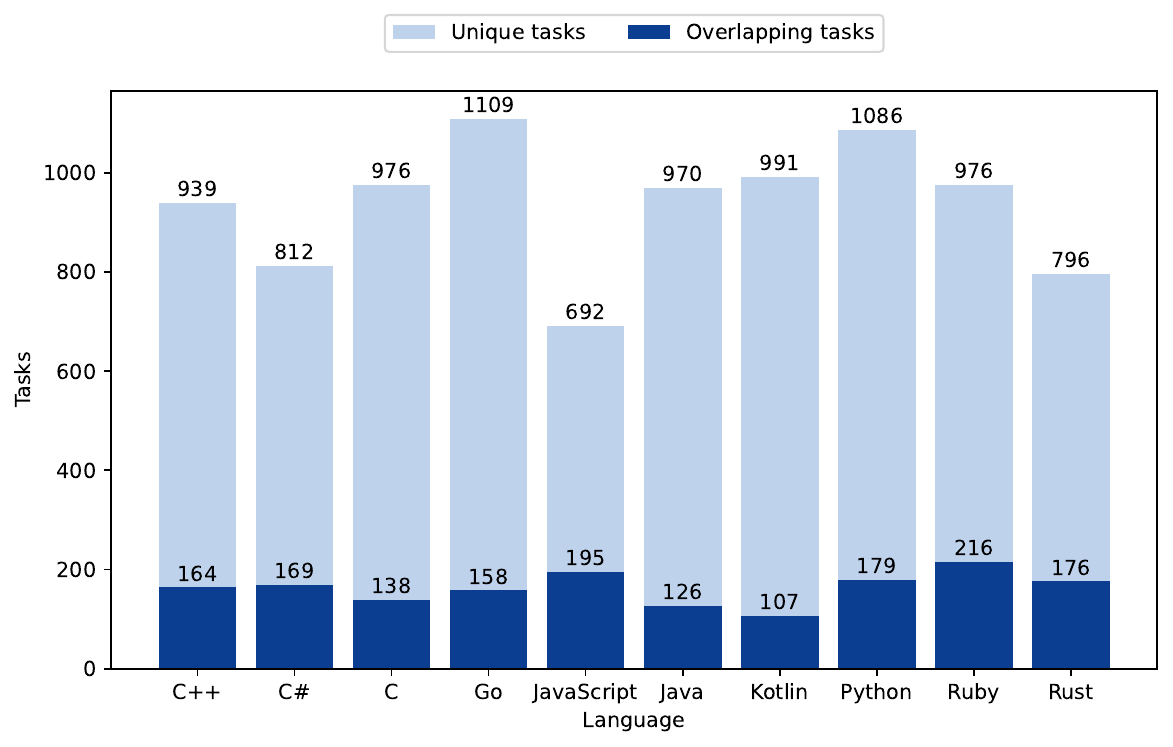}
  \caption{
    Distribution of unique tasks for which solutions are present in the dataset per language.
    For each task, both a human-written and an AI-written snippet is always provided.
    Overlapping tasks denote the number of tasks for which multiple AI-written solutions are present, with a guarantee that they have been translated from all \emph{other} programming languages in the dataset.
  }
  \label{fig:tasks}
\end{figure}

We recall from \Cref{sec:dataset-construction} that the generation of AI-labeled snippets in the dataset is followed by task balancing, ensuring that each language-specific sub-dataset contains \emph{pairs} of human-written/AI-written snippets solving the same task.
\Cref{fig:tasks} shows the number of unique tasks (or, equivalently, the number of snippet pairs) for which solutions are present in each sub-dataset, aggregated by programming language.
It also shows, for each language, the number of tasks for which there are solutions coming from all other provenance languages (``Overlapping tasks'' in the figure).

Due to the uneven distribution of task solutions across languages in Rosetta Code, different languages in the dataset show a different number of tasks present.
On the other hand, dataset users interested in avoiding the effect of translation provenance on AI-generated snippets can safely work in the subset of overlapping tasks.

\smallskip

In comparison to previous work in the literature~\cite{li_human_ai_styl, chatGPT_code_det_oedingen, hoqCS_course, codeDetectGPT_yang}, this dataset provides the ability to develop and test AI stylometry classifiers along multiple snippet dimensions---(1) programming language of the snippet, (2) provenance language (for AI-written snippets, obtained via translation), (3) snippet length, (4) task implemented by the snippet---allowing to isolate how each of them influences the performances of AI code stylometry.

\section{Results}
\label{sec:results}

We show the results of our experiments in two separate tables.
We first report, in \Cref{tab:multiPrompt}, accuracy results for all our models, both monolingual and multilingual.
Then we compare, in \Cref{tab:comp}, our best models to the best baselines from the literature.

We present all results in terms of overall accuracy, which is defined as the ratio of correctly classified snippets to the total number of snippets to be classified.
In order to establish the statistical significance of our results, we also conducted t-tests (see \Cref{tab:stochastic_results}) and ANOVA tests with $\alpha=0.05$ after testing data normality and variance.

\begin{table*}
  \caption{Accuracy (\%) of all trained classifiers when solving the AI code stylometry task.}
  \centering
  \label{tab:multiPrompt}
  \begin{tabular}{lcccccccccc|c} 
    \hline
    & \multicolumn{10}{c|}{\textbf{Language tested}}&\\
    \textbf{Prov. language} & \textbf{C++}  & \textbf{C} & \textbf{C\#} & \textbf{Go} & \textbf{Java}  & \textbf{Javascript}  & \textbf{Kotlin} & \textbf{Python} & \textbf{Ruby}  & \textbf{Rust} & \makecell{\textbf{Prov. accuracy} \\(avg. $\pm$ std)} \\  
    \hline \multicolumn{11}{c}{\emph{Monolingual models}} \\ \hline
    \textbf{C++ } & - & 88.3 & 87.8 & 91.0 & 81.4 & 86.2 & 91.0 & 87.2 & 86.2 & 84.0 & $87.0\pm2.9$ \\ 
    \textbf{C } & 89.9 & - & 90.9 & 91.5 & 92.0 & 90.9 & 94.1 & 89.9 & 88.8 & 88.8 & $90.8\pm2.0$\\ 
    \textbf{C\# } & 75.5 & 87.8 & - & 93.6 & 88.3 & 91.5 & 92.5 & 87.8 & 88.3 & 88.8 & $88.2\pm4.9$ \\ 
    \textbf{Go } & 94.7 & 94.7 & 87.2 & - & 90.9 & 95.2 & 94.7 & 84.0 & 90.4 & 85.6  & $90.8\pm4.1$ \\ 
    \textbf{Java } & 91.5 & 92.0 & 86.7 & 92.0 & - & 89.4 & 92.5 & 87.8 & 88.8 & 83.0 & $89.3\pm2.9$ \\ 
    \textbf{Javascript } & 90.9 & 95.2 & 88.3 & 91.5 & 87.7 & - & 93.1 & 85.6 & 90.4 & 84.0 & $89.6\pm3.4$\\ 
    \textbf{Kotlin } & 98.4 & 92.0 & 94.7 & 92.6 & 94.1 & 96.8 & - & 94.1 & 94.7 & 89.9 & \textbf{$94.1\pm2.4$} \\ 
    \textbf{Python } & 92.5 & 95.2 & 81.9 & 90.9 & 90.9 & 90.9 & 96.3 & - & 77.6 & 85.1 & $89.0\pm5.9$\\ 
    \textbf{Ruby } & 90.4 & 96.3 & 86.7 & 93.1 & 87.7 & 89.4 & 94.1 & 78.2 & - & 86.2 & $89.1\pm5.0$\\ 
    \textbf{Rust } & 88.3 & 97.8 & 90.9 & 87.3 & 94.7 & 91.5 & 87.8 & 88.8 & 89.4 & - & $90.7\pm3.3$\\ 
    \hline
    \textbf{Language accuracy} & $90.2$ & $93.2$ & $88.3$ & $91.5$ & $89.7$ & $91.3$ & $92.9$ & $87.0$ & $88.3$ & $86.1$ & - \\
    (avg. $\pm$ std) & $\pm$ & $\pm$ & $\pm$ & $\pm$ & $\pm$ & $\pm$ & $\pm$ & $\pm$ & $\pm$ & $\pm$ & \\
    & 5.9 & 3.3 & 3.4 & 1.7 & 3.8 & 3.0 & 2.3 & 4.1 & 4.3 & 2.3 & \\
    \hline \multicolumn{11}{c}{\emph{Multilingual model}} \\ \hline
    \textbf{Multilingual model accuracy}  & 88.8 & 88.3 & 84.0 & 89.4 & 82.4 & 79.3 & 87.8 & 79.3 & 80.8 & 81.4 & \textbf{84.1}$\,\pm\,$\textbf{3.8} \\ 
    \textbf{Multilingual model F1}  & 88.8 & 88.8 & 84.0 & 89.3 & 82.4 & 78.9 & 87.8 & 78.8 & 80.5 & 81.1 & \textbf{84.0}$\,\pm\,$\textbf{4.0} \\ 
    \textbf{Multilingual model AUC}  & 94.1 & 94.9 & 91.8 &  93.9 & 89.9 & 90.4 & 94.6 & 90.8 & 88.6 & 93.9  & \textbf{92.3}$\,\pm\,$\textbf{2.1} \\ 
    \hline
  \end{tabular}
\end{table*}

\begin{table}
  \caption{Hypothesis tests for \Cref{tab:multiPrompt}: ANOVA test results ($\alpha=0.05$) for Language accuracy; ANOVA test for  Provenance accuracy; t-test ($\alpha=0.05$) for Multilingual model and (monolingual) Language accuracy.}
  \centering
  \label{tab:stochastic_results}
  \begin{tabular}{lccc}
    \hline
    \textbf{Values}  & \textbf{t/F-statistic} &  \textbf{95\% CI} &\textbf{p-value}  \\  
    \hline
    \textbf{Lang. accuracy (ANOVA)}  &  3.56 & - & $<0.001$ \\ 
    \textbf{Prov. accuracy (ANOVA)} & 2.00 & - & 0.04 \\ 
    \textbf{Multilingual comp. (t-test)} & 4.01 & 2.7 to 8.7 &$<0.001$ \\ 
    \hline
  \end{tabular}
\end{table}

In \Cref{tab:multiPrompt}, we depict the results of the experiments across all languages and provenance.
Each monolingual classifier is tested only on snippets of the same language (in-distribution results), making the provenance language (i.e., the language AI-written snippets in the dataset were translated from) vary across all other languages. Therefore, we show the provenance language (\emph{src}) of the snippets in rows, while in the columns, we display the target language \emph{dst}, which is the language the models have been trained on.
The table also shows the average accuracy by provenance language (\emph{Prov. accuracy} column) and the average accuracy by tested language (\emph{Language accuracy} row).
The \emph{Multilingual model} row provides results for the multilingual classifier, which has been trained on the multilingual dataset sampled from \Dataset and tested separately on each language.
Finally, the bottom-right cell of \Cref{tab:multiPrompt} contains the average accuracy of the multilingual model across the 10 programming languages considered.

The results in \Cref{tab:multiPrompt} show significant differences both in rows (same provenance language and different destinations) and columns (different provenances and same destination).
\Cref{tab:stochastic_results} confirms that these differences among the models are significant, since we observed both for average provenance accuracies ($\mathrm{F{-}statistic}=2.00$ with $p=0.04$ in the table) and average language accuracy ($\mathrm{F{-}statistic}=3.56$ with $p<0.001$) relevant values.

\begin{table*}
  \caption{Comparison between the accuracy of baselines classifiers~\cite{li_human_ai_styl, chatGPT_code_det_oedingen} and classifiers introduced in this work.
    Best results for each column are shown in \textbf{bold}.
  }
  \label{tab:comp}
  \centering
  \begin{tabular}{lcccccccccc} 
    \hline
    &  \multicolumn{10}{c}{\bfseries Tested language} \\
    \textbf{Model} &  \textbf{Java}  & \textbf{C++} & \textbf{Python} & \textbf{Javascript} & \textbf{C\#}  & \textbf{C}  & \textbf{Go}  & \textbf{Ruby}  & \textbf{Rust}  & \textbf{Kotlin}  \\  
    \hline \multicolumn{11}{c}{\emph{Baseline models}} \\ \hline
    \textbf{RF Java~\cite{li_human_ai_styl}} & 79.3 & - & -& -& -& -& -& -& -& - \\ 
    \textbf{J48 Java~\cite{li_human_ai_styl}} &  86.9 &  - & -& -& -& -& -& -& -& - \\ 
    \textbf{RF C++~\cite{li_human_ai_styl}} & -  & 85.5 & -& -& -& -& -& -& -& - \\
    \textbf{J48 C++~\cite{li_human_ai_styl}} &  - &  89.5 & -& -& -& -& -& -& -& - \\ 
    \textbf{XGB-TF-IDF Python~\cite{chatGPT_code_det_oedingen}} &  - &  - & 92.6 & -& -& -& -& -& -& - \\ 
    \hline \multicolumn{11}{c}{\emph{Proposed models}} \\ \hline
    \textbf{Java} (monolingual) &  \textbf{94.1} & 83.9 & 87.7  & 90.6 & \textbf{88.9} & 80.4 & 86.7 & 86.6& 85.0& 31.3   \\ 
    \textbf{C++} (monolingual) & 89.3 & \textbf{98.4} & 88.0 & \textbf{90.7} & 86.7 & \textbf{89.5} & 86.1 & 87.0 & \textbf{86.2} & 24.1  \\ 
    \textbf{Python} (monolingual) & 63.1 & 82.4 & \textbf{94.1} & 90.2 &58.5 & 83.6 & 58.7 & \textbf{90.7} &  84.0 & 44.1  \\ 
    \textbf{Multilingual}  & 82.4 &88.8 & 79.3 & 79.3 & 84.0 & 88.3 &  \textbf{89.4} &80.8  & 81.4 & \textbf{87.8}  \\ 
    \hline
  \end{tabular}
\end{table*}

In \Cref{tab:comp} we compare our models with the baseline ones, namely J48 and Random forest algorithms from Li et al.~\cite{li_human_ai_styl} and XGB from Oedingen et al.~\cite{chatGPT_code_det_oedingen}.
All the baselines and our monolingual models (Java, C++, Python) are trained with our best dataset, namely the one that provides the best value for column \emph{Prov.~accuracy} in \Cref{tab:multiPrompt}, that is Kotlin. 
The test datasets are also all with Kotlin provenance, except for Kotlin itself which has Go provenance.

For monolingual models---when analyzing in-distribution tests---we highlight a substantial positive gap (+7.2\% for the Java language, +8.9\% for the C++ language, and +1.5\% for the Python model) compared to the baselines.
We notice the positive gap between the C++ model tested on the Java (out-distribution test) dataset (+2.4\%), showing how, even when trained on a different language, this methodology performs better than the one adopted by Li et al.~\cite{li_human_ai_styl}.
Out-distribution tests for the baselines are not shown as these architectures employ predefined features---lexical, syntactical (extracted from the abstract syntax tree of the source code), or deriving from source code layout---strictly linked to the language used during training, making the model specific to the designed language (a net advantage for the approach proposed in this paper).

We notice how the multilingual classifier performs worse than monolingual classifiers in both \Cref{tab:multiPrompt} and \Cref{tab:comp}. In particular, \Cref{tab:stochastic_results} shows that the multilingual classifier has an accuracy that is worse than that one of the monolingual model in the average language case (-5.17\% avg. $t{-}statistic=4.01$ with $p<0.001$).
The multilingual classifier \Cref{tab:comp}, however, does not present outliers in terms of accuracies, obtaining a model with consistent results, effective in handling multiple languages and the provenance phenomenon.
In addition to the practical benefits of having a single model, this is another reason, as we will discuss in \Cref{sec:discussion}, why the multilingual model is preferable for detecting AI-generated code in practice.

We also observe that our reimplementations of the baselines perform worse than the results reported in the original papers by both Li et al.~\cite{li_human_ai_styl} and Oedingen et al.~\cite{chatGPT_code_det_oedingen}.
Specifically, we obtained negative gaps of -18.5\% with Java~\cite{li_human_ai_styl}, -3.5\% with the C++ baseline~\cite{li_human_ai_styl} and -5.2\% for the Python baseline~\cite{chatGPT_code_det_oedingen}.
Since we reimplemented the same methodologies and applied them on our dataset (to perform a fair comparison on an identical dataset), we attribute these differences to the dataset itself, suggesting that \Dataset is a harder benchmark than the ones used in previous work---which also makes intuitive sense, given the fast-paced advances in LLM-based code generators.

\section{Discussion}
\label{sec:discussion}

\paragraph{Findings}

Based on the results presented in \Cref{sec:results} we can answer the stated research question affirmatively: it is possible to recognize AI-written programs with high average accuracy (\DataAvgAccuracy), across 10 different programming language (multilingual code stylometry), with a single trained classifier based on a transformer-based architecture, novel for this task.

To achieve this, we devised and implemented a fully open and reproducible methodology and also replicated previous experiments in the literature~\cite{li_human_ai_styl, chatGPT_code_det_oedingen}.
We observe significant performance differences not only across different datasets (which is to be expected), but also between accuracies previously reported in the literature and our replications of the same experiments with the same architectures.
The following factors might be the cause of these discrepancies:

(1) The language in which the snippets to be recognized are written in plays an important role.
  For example, we see a pattern in our results comparing C snippets (accuracy $93.2\%\pm3.3$) with Rust snippets ($86.1\%\pm2.3$), for a difference of 7 points.
  It is entirely possible that AI-written code in some programming languages is intrinsically easier to recognize as such than code written in other languages.
  Establishing this conclusively is an interesting direction for future work.
\begin{figure}
  \centering
  \includegraphics[width=\linewidth]{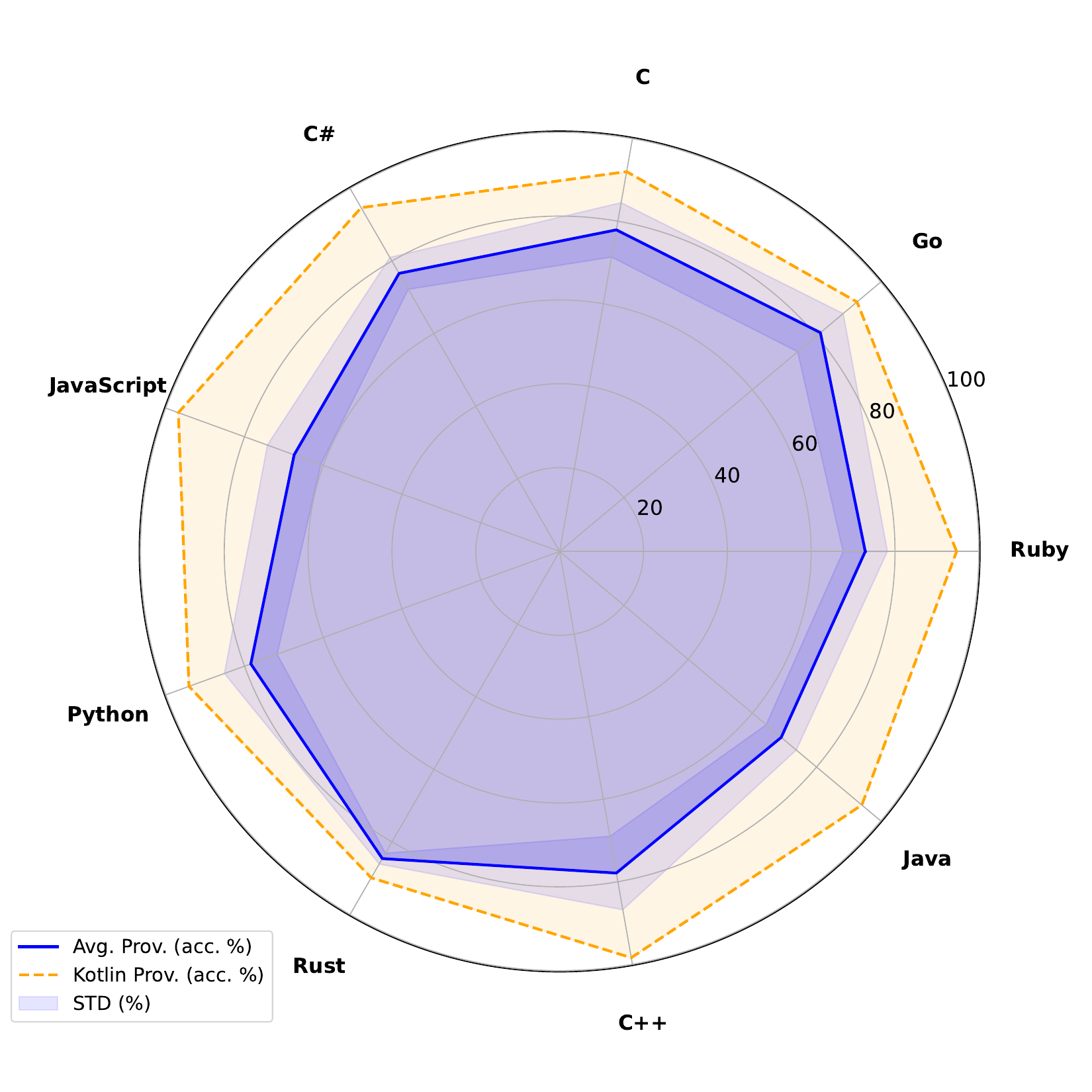}
  \caption{Accuracy of our best-performing monolingual models.
    The dashed orange line corresponds to the Kotlin row in \Cref{tab:multiPrompt} (in-distribution).
    The solid blue line shows the accuracy of the same models on test sets with a different language provenance (potentially out-distribution, as is general translating solutions for the same task from different languages to the same language will result in different snippets).
    Model performances degrade significantly in the latter case.
  }
  \label{fig:prov}
\end{figure}

(2) We also observe significant accuracy differences between datasets with different provenance languages (i.e., the languages AI-written code snippets where translated \emph{from}).
In order to test the impact of this factor, we analyzed models in the same language but with different provenances for the testing datasets (\Cref{fig:prov}).
More precisely, since for the same language we have different models (depending on the provenance language), we first selected the models with the best provenance accuracy (last column of \Cref{tab:multiPrompt}), namely the models with Kotlin provenance.
We depicted in \Cref{fig:prov}, with the dashed orange line, the accuracy of these models on the various languages (that is, we reported in the Kotlin row of \Cref{tab:multiPrompt}).
Then, we compared these accuracies to those obtained for each language $L \neq \mathrm{Kotlin}$ by averaging the accuracies of the $L$ model on all the training datasets generated in $L$ from the other provenances.
This is reported in \Cref{fig:prov} with the blue line.

As the picture shows, we obtain a significant loss in accuracy (-17.93\% in avg. $\mathrm{t{-}statistic}=8.82$ with 95\% CI 12.4 to 21.3 and  $p<0.001$). 
We believe this phenomenon is tightly coupled with the methodological choice of producing AI-written snippets via code translation and would not be observable under different conditions.
We consider that we have properly mitigated this by training our main classifier to be multilingual across multiple translation provenances.

Still, the accuracy differences are interesting \emph{per se} and pave the way to dedicated future work.
We hypothesize that different ways of inputting---and prompting---the generative model, both to generate the training datasets and to use the obtained classifiers, can influence its performance, possibly leading to detection evasion.

(3) Reproducibility issues.
  As is way too common in empirical software engineering~\cite{barahona2023swe-rep}, studies tend to under-report the details needed to fully replicate their empirical findings, particularly so when AI-training is involved.
  When replicating baseline results from previous work, we tried our best, but some artifacts were not available and had to be reimplemented from scratch.
  This work contributes to raising the bar of replicability for AI code stylometry by relying on and producing only openly available artifacts.

\smallskip
Aside from the above, in \Cref{sec:dataset}, we highlighted two other factors that can influence detection accuracy: dataset size and task balancing.
We did not explore systematically how either of them influence the results; it is left as future work.

\smallskip
\paragraph{What about ChatGPT?}
\label{sec:chatgpt}

ChatGPT is one of the most popular generative AI tools on the market, for both generating natural language text and program code.
We could not use ChatGPT in our experiments without undermining their replicability---which is one of our goals and differentiating aspects w.r.t.~previous work.
Still, it is interesting to verify how our reproducible classifiers, trained using only openly available data, fare against ChatGPT (whose model and training data are undisclosed).

We tested our multilingual classifier (trained on the \Dataset dataset) on the Human-AI dataset by Oedingen et al.~\cite{chatGPT_code_det_oedingen}, whose AI part consists of Python code snippets generated by ChatGPT (at the time and version of their experiments) starting from natural language prompts.
Without any fine-tuning on ChatGPT data, our multilingual classifier achieved 72.8\% accuracy, with a -6.3\% drop from the results obtained on our dataset.
The accuracy gap is more significant when compared to the results reported by Oedingen et al.~\cite{chatGPT_code_det_oedingen}: -25.2\% with our model on their dataset.
Still, our classifier results are way above the chance (50\% for 2 classes), multilingual, and reproducible.

The accuracy gap can be due to several factors: the LLM used to generate the AI snippets, the different inputs and prompts used during generation, hyperparameter tuning, etc.
Ultimately, the important question here is whether, in the upcoming arm race between AI code generation and AI code stylometry (to detect it), we can rely on closed models and datasets or not.
We argue we should not and propose a new baseline for reproducible, multilingual, AI code stylometry with this work.

\section{Limitations}
\label{sec:threats}

\paragraph*{External validity}
Our approach to the generation of AI-written snippets in the novel \Dataset dataset is reproducible code translation.
Our results could be impacted by that choice.
We have explored the possibility only superficially, by verifying how our classifier performs on the dataset of Oedingen et al.~\cite{chatGPT_code_det_oedingen}, generated using ChatGPT on natural language prompts (hence: not code translation), with good results (cf.~\Cref{sec:discussion}).
A more thorough analysis of how our results generalize to other prompting techniques (e.g., natural language prompts) and LLMs is left as future work.
Note that we share this threat with all related work and that it is impossible to fully mitigate this threat encompassing closed LLMs (like ChatGPT) without sacrificing reproducibility.

\paragraph*{Reliability}
We address reliability threats in the usual way by releasing a comprehensive replication package covering all experiments discussed in the paper (see the Data availability statement at the end).
In this respect, we fare better than all previous works by relying only on openly available datasets and components, including third-party LLMs.

\section{Related work}
\label{sec:related}

\paragraph{Code stylometry}

Oman and Cook~\cite{oman1989FirstStyl} were the first to introduce the notion of code stylometry.
They hypothesized that each author is recognizable by a unique coding style called ``fingerprint''.
In their pioneering work, they approached the task using cluster-based classification, introducing an unsupervised technique for inferring the code author.

Caliskan-Islam et al.~\cite{caliskan2015stylAST} were the first to use both syntactical features (from ASTs) and lexical features (from concrete syntax trees) for code stylometry.
They showed how a random forest classifier can take advantage of both kinds of information to achieve an accuracy of 53.91\% for the Python language across 229 different authors.
Other works followed the Caliskan approach, e.g., Dauber et al.~\cite{dauber2017GitHubStyle}.

More recently, the emergence of word embeddings~\cite{mikolov2013wordEmbeddings} introduced a shift in the representation of author fingerprints from classical machine-learning techniques to deep-learning ones.
Deep learning approaches~\cite{alsulami2017LSTMstyl, bogomolov, kovalenko2020codeStylEmb} led to better author style representation capabilities, most notably by leveraging LSTM and code2vec~\cite{alon2018code2vec} architectures.
In terms of code stylometry accuracy, this resulted in a bump up to 95.90\% with 70 different authors~\cite{bogomolov}.
These architectures represent source code using both syntactical and lexical features.

Our work in this paper is a specific instance of the code stylometry task, where we aim to distinguish a specific ``AI author'' (a code LLM) from human authors.
To that end we introduce the use of a transformer-based architecture~\cite{attentionIsAllYouNeed}, novel for the code stylometry task.
Contrary to more traditional code stylometry work, we do not rely on syntactical features, but solely on lexical features (token stream).

\paragraph{AI detection for natural language}

Köbis and Mossink~\cite{unrecognizable_AI} observed first how the generative capabilities of LLMs make it difficult to distinguish their (natural language, in this case) output from human-authored text, paving the way to research on the topic.

Early studies~\cite{real_fake_bakhtin2019, real_fake_harada2021, real_fake_gehrmann2019gltr} approached the problem of AI-generated natural language using stochastic approaches.
They generated AI-labeled samples with GPT-2~\cite{radford2019language} and reached accuracies up to 93\%~\cite{real_fake_bakhtin2019}.

Liao et al.~\cite{liaoChatGPT} introduced the use of BERT~\cite{BERT} for recognizing AI-generated natural language.
They demonstrated that this approach results in superior accuracy (96.7\%) compared to traditional machine learning approaches: +7\% w.r.t.~XGBoost (decision tree) to a fine-tuned BERT architecture.

Mitchell et al.~\cite{mitchell2023detectgpt} used an approach based solely on probabilities sampled from a generative model, reaching 86\% AUROC.
Mitrović et al.~\cite{mitrovic2023chatgpt} compared the performance of different approaches that do not need fine-tuning.
They show that supervised techniques perform better, with a 14\% accuracy increase from a perplexity-based approach (84\%) to DistillBERT (98\%).

With respect to these works, we focus on AI \emph{code} stylometry, rather than natural language.
We adopt an LLM-based approach (like~\cite{liaoChatGPT, mitrovic2023chatgpt}), fine tuning the T5plus~\cite{wang2023codet5+} LLM.
For dataset generation, we use the open-weight StarCoder2~\cite{starcoder2} code LLM.

\paragraph{AI code stylometry}

Hoq et al.~\cite{hoqCS_course} looked at the problem of AI \emph{code} stylometry, for educational plagiarism detection in the context of university computer science class.
Their dataset consists of: (1) student-written Java code from a publicly-available dataset encompassing multiple problems with human solutions, and (2) solutions to the same problems generated by ChatGPT.
Their approach relied on both syntactical and lexical features extracted from the code, fed to both traditional machine learning techniques (random forest) and deep learning ones, such as code2vec~\cite{alon2018code2vec}.
They reached accuracies up to 95\% (with code2vec).

Both Bukhari et al.~\cite{bukariFirst} and Idialu et al.~\cite{Human_AI_python} followed the same approach on different datasets and programming languages.
The former focused on the C language using the Lost at C dataset~\cite{sandoval2023code_llm_security} and Codex~\cite{codex} for AI-generated snippets.
The latter looked at Python code from the CodeChef learning platform, and generated the AI solutions with GPT-4.
Both studies used random forest classification, reaching respectively 92\% accuracy and 91\% F1-score.

Yang et al.~\cite{codeDetectGPT_yang} replicated for AI code stylometry the probability-based methodology introduced by Mitchell et al.~\cite{mitchell2023detectgpt} for natural language.
They focused on Java and Python, data coming from multiple LLMs from OpenAI, reaching AUCs up to 86.01\% for Python and 77.42\% for Java.

Oedingen et al.~\cite{chatGPT_code_det_oedingen} analyzed the discrepancies between fine-tuned methodologies and zero-shot approaches (like Yang et al.~\cite{codeDetectGPT_yang}), showing how the latter struggle to achieve competitive performances.
Using traditional machine learning techniques (XGB with TF-IDF features), they achieved impressive results (98\% accuracy) on the detection of Python code generated by ChatGPT.

Rahman et al.~\cite{claude_detect} followed a similar approach, using a different (but still proprietary) LLM for code generation: Claude 3 haiku~\cite{claude3}.
They reached 82\% accuracy on Python.

Li et al.~\cite{li_human_ai_styl} introduced the use of translations (from either natural language specifications or existing code) to generate the AI-labeled part of training datasets for AI code stylometry.
Using translation, they aim to reduce the chance of producing code already present in the training dataset of the code LLM that is to be recognized as an AI author.
They considered C++ and Java languages (\emph{separately}), using both ChatGPT and GPT-4 as generative models.
They used random forest classification, with only lexical features, achieving 93\% and 97.8\% accuracy for C++ and Java, respectively.

We depart from previous AI code stylometry work in three ways: (1) we apply for the first time a transformer architecture to the AI code stylometry task; (2) we are able to recognize AI-written code across 10 different programming languages with a single model, achieving an average accuracy of \DataAvgAccuracy; (3) we rely only on openly-available data and code, enabling scientific reproducibility and future reuse of our work.

\paragraph{GPTSniffer}

Independently and in parallel to our work, Nguyen et al.~\cite{GPTSniffer} introduced GPT Sniffer, that tackles the task of detecting AI-generated coming from a different perspective than code stylometry, but also using a transformer-based classifier.
Their work is focused on the Java language, uses ChatGPT as a generative model, and considers the impact of different data source domains, such as programming books and data representative of real use-case scenarios that encompass a mixture of snippets from GitHub repositories and generated from ad-hoc queries.
GPTSniffer performs really well in domain (100\% F1 score training and testing with data provenance from a Java programming book), degrades a lot on out-of-domain (56\% F1 with real use-case data for testing), and improves again on a mixed training set (94\% F1) and data alteration techniques (96\% F1). 
In comparison to GPTSniffer, we rely exclusively on open models for data generation and classification, ensuring experiment reproducibility, and supports a larger classification scope of 10 distinct programming languages with a single model.

\section{Conclusions}
\label{sec:conclusions}

AI code stylometry consists of automatically detecting whether an input piece of source code was authored by an AI (e.g., Copilot or ChatGPT) or a human.
In this paper, we took a fresh look at the problem by revisiting assumptions made in previous work.

First, we solved the problem in a multilingual setting, supporting 10 different popular programming languages achieving high average accuracy (\DataAvgAccuracy) with a single transformed-based classifier, a novel architecture for this task.

Second, our experiments are fully reproducible.
As building blocks, we use only openly available data and components, including our code LLM: StarCoder2.
We release openly all our artifacts: the novel \Dataset dataset consisting of \DataSnippetsMp code snippets in 10 languages, partly human-written (from Rosetta Code) and partly AI-written (via cross-language code translation); checkpoint of our trained model; and an open source CLI tool to use it in practice.

As future work we plan to analyze how snippet generation impacts detection accuracy, covering: prompt engineering, used code LLM, the provenance language for code translation, as well as starting from natural language prompts.

\section*{Data availability} 
A full replication package containing the dataset, a checkpoint of our multilingual model, a command-line tool for using it on selected code snippets, as well as the source code used to run all the experiments presented in this paper is available from Zenodo at \url{https://doi.org/10.5281/zenodo.13908858}.


\begin{thebibliography}{10}

\bibitem{agarwal2024copilot_eval}
Anisha Agarwal, Aaron Chan, Shubham Chandel, Jinu Jang, Shaun Miller,
  Roshanak~Zilouchian Moghaddam, Yevhen Mohylevskyy, Neel Sundaresan, and
  Michele Tufano.
\newblock Copilot evaluation harness: Evaluating llm-guided software
  programming.
\newblock {\em CoRR}, abs/2402.14261, 2024.

\bibitem{alon2018code2vec}
Uri Alon, Meital Zilberstein, Omer Levy, and Eran Yahav.
\newblock code2vec: learning distributed representations of code.
\newblock {\em Proc. {ACM} Program. Lang.}, 3({POPL}):40:1--40:29, 2019.

\bibitem{alsulami2017LSTMstyl}
Bander Alsulami, Edwin Dauber, Richard~E. Harang, Spiros Mancoridis, and Rachel
  Greenstadt.
\newblock Source code authorship attribution using long short-term memory based
  networks.
\newblock In Simon~N. Foley, Dieter Gollmann, and Einar Snekkenes, editors,
  {\em Computer Security - {ESORICS} 2017 - 22nd European Symposium on Research
  in Computer Security, Oslo, Norway, September 11-15, 2017, Proceedings, Part
  {I}}, volume 10492 of {\em Lecture Notes in Computer Science}, pages 65--82.
  Springer, 2017.

\bibitem{claude3}
Anthropic.
\newblock The claude 3 model family: Opus, sonnet, haiku.
\newblock https://www.anthropic.com/claude-3-model-card, 2024.

\bibitem{real_fake_bakhtin2019}
Anton Bakhtin, Sam Gross, Myle Ott, Yuntian Deng, Marc'Aurelio Ranzato, and
  Arthur Szlam.
\newblock Real or fake? learning to discriminate machine from human generated
  text.
\newblock {\em CoRR}, abs/1906.03351, 2019.

\bibitem{bogomolov}
Egor Bogomolov, Vladimir Kovalenko, Yurii Rebryk, Alberto Bacchelli, and
  Timofey Bryksin.
\newblock Authorship attribution of source code: a language-agnostic approach
  and applicability in software engineering.
\newblock In Diomidis Spinellis, Georgios Gousios, Marsha Chechik, and
  Massimiliano~Di Penta, editors, {\em {ESEC/FSE} '21: 29th {ACM} Joint
  European Software Engineering Conference and Symposium on the Foundations of
  Software Engineering, Athens, Greece, August 23-28, 2021}, pages 932--944.
  {ACM}, 2021.

\bibitem{bukariFirst}
Sufiyan Bukhari, Benjamin Tan, and Lorenzo~De Carli.
\newblock Distinguishing {AI-} and human-generated code: {A} case study.
\newblock In Santiago Torres{-}Arias, Marcela~S. Melara, Laurent Simon, Nikos
  Vasilakis, and Kathleen Moriarty, editors, {\em Proceedings of the 2023
  Workshop on Software Supply Chain Offensive Research and Ecosystem Defenses,
  {SCORED} 2023, Copenhagen, Denmark, 30 November 2023}, pages 17--25. {ACM},
  2023.

\bibitem{codex}
Mark Chen, Jerry Tworek, Heewoo Jun, Qiming Yuan, Henrique~Pond{\'{e}}
  de~Oliveira~Pinto, Jared Kaplan, Harri Edwards, Yuri Burda, Nicholas Joseph,
  Greg Brockman, Alex Ray, Raul Puri, Gretchen Krueger, Michael Petrov, Heidy
  Khlaaf, Girish Sastry, Pamela Mishkin, Brooke Chan, Scott Gray, Nick Ryder,
  Mikhail Pavlov, Alethea Power, Lukasz Kaiser, Mohammad Bavarian, Clemens
  Winter, Philippe Tillet, Felipe~Petroski Such, Dave Cummings, Matthias
  Plappert, Fotios Chantzis, Elizabeth Barnes, Ariel Herbert{-}Voss,
  William~Hebgen Guss, Alex Nichol, Alex Paino, Nikolas Tezak, Jie Tang, Igor
  Babuschkin, Suchir Balaji, Shantanu Jain, William Saunders, Christopher
  Hesse, Andrew~N. Carr, Jan Leike, Joshua Achiam, Vedant Misra, Evan Morikawa,
  Alec Radford, Matthew Knight, Miles Brundage, Mira Murati, Katie Mayer, Peter
  Welinder, Bob McGrew, Dario Amodei, Sam McCandlish, Ilya Sutskever, and
  Wojciech Zaremba.
\newblock Evaluating large language models trained on code.
\newblock {\em CoRR}, abs/2107.03374, 2021.

\bibitem{rosetta-code}
Rosetta Code.
\newblock Rosetta code.
\newblock https://rosettacode.org, Jul 2022.
\newblock [Online; accessed 1-July-2022].

\bibitem{softwareHeritage}
Roberto~Di Cosmo and Stefano Zacchiroli.
\newblock Software heritage: Why and how to preserve software source code.
\newblock In Shoichiro Hara, Shigeo Sugimoto, and Makoto Goto, editors, {\em
  Proceedings of the 14th International Conference on Digital Preservation,
  iPRES 2017, Kyoto, Japan, September 25-29, 2017}, 2017.

\bibitem{dakhel2023copilot_eval}
Arghavan~Moradi Dakhel, Vahid Majdinasab, Amin Nikanjam, Foutse Khomh,
  Michel~C. Desmarais, and Zhen Ming~(Jack) Jiang.
\newblock {GitHub} {Copilot} {AI} pair programmer: Asset or liability?
\newblock {\em J. Syst. Softw.}, 203:111734, 2023.

\bibitem{dauber2017GitHubStyle}
Edwin Dauber, Aylin Caliskan, Richard~E. Harang, Gregory Shearer, Michael~J.
  Weisman, Frederica Free{-}Nelson, and Rachel Greenstadt.
\newblock Git blame who?: Stylistic authorship attribution of small, incomplete
  source code fragments.
\newblock {\em Proc. Priv. Enhancing Technol.}, 2019(3):389--408, 2019.

\bibitem{BERT}
Jacob Devlin, Ming{-}Wei Chang, Kenton Lee, and Kristina Toutanova.
\newblock {BERT:} pre-training of deep bidirectional transformers for language
  understanding.
\newblock In Jill Burstein, Christy Doran, and Thamar Solorio, editors, {\em
  Proceedings of the 2019 Conference of the North American Chapter of the
  Association for Computational Linguistics: Human Language Technologies,
  {NAACL-HLT} 2019, Minneapolis, MN, USA, June 2-7, 2019, Volume 1 (Long and
  Short Papers)}, pages 4171--4186. Association for Computational Linguistics,
  2019.

\bibitem{real_fake_gehrmann2019gltr}
Sebastian Gehrmann, Hendrik Strobelt, and Alexander~M. Rush.
\newblock {GLTR:} statistical detection and visualization of generated text.
\newblock pages 111--116, 2019.

\bibitem{barahona2023swe-rep}
Jes{\'{u}}s~M. Gonz{\'{a}}lez{-}Barahona and Gregorio Robles.
\newblock Revisiting the reproducibility of empirical software engineering
  studies based on data retrieved from development repositories.
\newblock {\em Inf. Softw. Technol.}, 164:107318, 2023.

\bibitem{deepseek}
Daya Guo, Qihao Zhu, Dejian Yang, Zhenda Xie, Kai Dong, Wentao Zhang, Guanting
  Chen, Xiao Bi, Y.~Wu, Y.~K. Li, Fuli Luo, Yingfei Xiong, and Wenfeng Liang.
\newblock Deepseek-coder: When the large language model meets programming - the
  rise of code intelligence.
\newblock {\em CoRR}, abs/2401.14196, 2024.

\bibitem{real_fake_harada2021}
Atsumu Harada, Danushka Bollegala, and Naiwala~P. Chandrasiri.
\newblock Discrimination of human-written and human and machine written
  sentences using text consistency.
\newblock In {\em 2021 International Conference on Computing, Communication,
  and Intelligent Systems (ICCCIS)}, pages 41--47, 2021.

\bibitem{hoqCS_course}
Muntasir Hoq, Yang Shi, Juho Leinonen, Damilola Babalola, Collin~F. Lynch, and
  Bita Akram.
\newblock Detecting chatgpt-generated code in a {CS1} course.
\newblock In Steven Moore, John~C. Stamper, Richard~Jiarui Tong, Chen Cao,
  Zitao Liu, Xiangen Hu, Yu~Lu, Joleen Liang, Hassan Khosravi, Paul Denny,
  Anjali Singh, and Christopher Brooks, editors, {\em Proceedings of the
  Workshop on Empowering Education with LLMs - the Next-Gen Interface and
  Content Generation 2023 co-located with 24th International Conference on
  Artificial Intelligence in Education {(AIED} 2023), Tokyo, Japan, July 7,
  2023}, volume 3487 of {\em {CEUR} Workshop Proceedings}, pages 53--63.
  CEUR-WS.org, 2023.

\bibitem{Human_AI_python}
Oseremen~Joy Idialu, Noble~Saji Mathews, Rungroj Maipradit, Joanne~M. Atlee,
  and Meiyappan Nagappan.
\newblock Whodunit: Classifying code as human authored or {GPT-4} generated-
  {A} case study on codechef problems.
\newblock In Diomidis Spinellis, Alberto Bacchelli, and Eleni Constantinou,
  editors, {\em 21st {IEEE/ACM} International Conference on Mining Software
  Repositories, {MSR} 2024, Lisbon, Portugal, April 15-16, 2024}, pages
  394--406. {ACM}, 2024.

\bibitem{caliskan2015stylAST}
Aylin~Caliskan Islam, Richard~E. Harang, Andrew Liu, Arvind Narayanan, Clare~R.
  Voss, Fabian Yamaguchi, and Rachel Greenstadt.
\newblock De-anonymizing programmers via code stylometry.
\newblock In Jaeyeon Jung and Thorsten Holz, editors, {\em 24th {USENIX}
  Security Symposium, {USENIX} Security 15, Washington, D.C., USA, August
  12-14, 2015}, pages 255--270. {USENIX} Association, 2015.

\bibitem{unrecognizable_AI}
Nils~C. K{\"{o}}bis and Luca Mossink.
\newblock Artificial intelligence versus maya angelou: Experimental evidence
  that people cannot differentiate ai-generated from human-written poetry.
\newblock {\em Comput. Hum. Behav.}, 114:106553, 2021.

\bibitem{kovalenko2020codeStylEmb}
Vladimir Kovalenko, Egor Bogomolov, Timofey Bryksin, and Alberto Bacchelli.
\newblock Building implicit vector representations of individual coding style.
\newblock In {\em {ICSE} '20: 42nd International Conference on Software
  Engineering, Workshops, Seoul, Republic of Korea, 27 June - 19 July, 2020},
  pages 117--124. {ACM}, 2020.

\bibitem{li_human_ai_styl}
Ke~Li, Sheng Hong, Cai Fu, Yunhe Zhang, and Ming Liu.
\newblock Discriminating human-authored from chatgpt-generated code via
  discernable feature analysis.
\newblock In {\em 34th {IEEE} International Symposium on Software Reliability
  Engineering, {ISSRE} 2023 - Workshops, Florence, Italy, October 9-12, 2023},
  pages 120--127. {IEEE}, 2023.

\bibitem{starcoder}
Raymond Li, Loubna~Ben Allal, Yangtian Zi, Niklas Muennighoff, Denis Kocetkov,
  Chenghao Mou, Marc Marone, Christopher Akiki, Jia Li, Jenny Chim, Qian Liu,
  Evgenii Zheltonozhskii, Terry~Yue Zhuo, Thomas Wang, Olivier Dehaene, Mishig
  Davaadorj, Joel Lamy{-}Poirier, Jo{\~{a}}o Monteiro, Oleh Shliazhko, Nicolas
  Gontier, Nicholas Meade, Armel Zebaze, Ming{-}Ho Yee, Logesh~Kumar Umapathi,
  Jian Zhu, Benjamin Lipkin, Muhtasham Oblokulov, Zhiruo Wang, Rudra~Murthy V,
  Jason~T. Stillerman, Siva~Sankalp Patel, Dmitry Abulkhanov, Marco Zocca,
  Manan Dey, Zhihan Zhang, Nour Fahmy, Urvashi Bhattacharyya, Wenhao Yu, Swayam
  Singh, Sasha Luccioni, Paulo Villegas, Maxim Kunakov, Fedor Zhdanov, Manuel
  Romero, Tony Lee, Nadav Timor, Jennifer Ding, Claire Schlesinger, Hailey
  Schoelkopf, Jan Ebert, Tri Dao, Mayank Mishra, Alex Gu, Jennifer Robinson,
  Carolyn~Jane Anderson, Brendan Dolan{-}Gavitt, Danish Contractor, Siva Reddy,
  Daniel Fried, Dzmitry Bahdanau, Yacine Jernite, Carlos~Mu{\~{n}}oz Ferrandis,
  Sean Hughes, Thomas Wolf, Arjun Guha, Leandro von Werra, and Harm de~Vries.
\newblock Starcoder: may the source be with you!
\newblock {\em Trans. Mach. Learn. Res.}, 2023, 2023.

\bibitem{liaoChatGPT}
Wenxiong Liao, Zhengliang Liu, Haixing Dai, Shaochen Xu, Zihao Wu, Yiyang
  Zhang, Xiaoke Huang, Dajiang Zhu, Hongmin Cai, Tianming Liu, and Xiang Li.
\newblock Differentiate chatgpt-generated and human-written medical texts.
\newblock {\em CoRR}, abs/2304.11567, 2023.

\bibitem{AdamW}
Ilya Loshchilov and Frank Hutter.
\newblock Decoupled weight decay regularization.
\newblock In {\em 7th International Conference on Learning Representations,
  {ICLR} 2019, New Orleans, LA, USA, May 6-9, 2019}. OpenReview.net, 2019.

\bibitem{starcoder2}
Anton Lozhkov, Raymond Li, Loubna~Ben Allal, Federico Cassano, Joel
  Lamy{-}Poirier, Nouamane Tazi, Ao~Tang, Dmytro Pykhtar, Jiawei Liu, Yuxiang
  Wei, Tianyang Liu, Max Tian, Denis Kocetkov, Arthur Zucker, Younes Belkada,
  Zijian Wang, Qian Liu, Dmitry Abulkhanov, Indraneil Paul, Zhuang Li,
  Wen{-}Ding Li, Megan Risdal, Jia Li, Jian Zhu, Terry~Yue Zhuo, Evgenii
  Zheltonozhskii, Nii Osae~Osae Dade, Wenhao Yu, Lucas Krau{\ss}, Naman Jain,
  Yixuan Su, Xuanli He, Manan Dey, Edoardo Abati, Yekun Chai, Niklas
  Muennighoff, Xiangru Tang, Muhtasham Oblokulov, Christopher Akiki, Marc
  Marone, Chenghao Mou, Mayank Mishra, Alex Gu, Binyuan Hui, Tri Dao, Armel
  Zebaze, Olivier Dehaene, Nicolas Patry, Canwen Xu, Julian~J. McAuley, Han Hu,
  Torsten Scholak, S{\'{e}}bastien Paquet, Jennifer Robinson, Carolyn~Jane
  Anderson, Nicolas Chapados, and et~al.
\newblock Starcoder 2 and the stack v2: The next generation.
\newblock {\em CoRR}, abs/2402.19173, 2024.

\bibitem{mikolov2013wordEmbeddings}
Tom{\'{a}}s Mikolov, Kai Chen, Greg Corrado, and Jeffrey Dean.
\newblock Efficient estimation of word representations in vector space.
\newblock In Yoshua Bengio and Yann LeCun, editors, {\em 1st International
  Conference on Learning Representations, {ICLR} 2013, Scottsdale, Arizona,
  USA, May 2-4, 2013, Workshop Track Proceedings}, 2013.

\bibitem{mitchell2023detectgpt}
Eric Mitchell, Yoonho Lee, Alexander Khazatsky, Christopher~D. Manning, and
  Chelsea Finn.
\newblock Detectgpt: Zero-shot machine-generated text detection using
  probability curvature.
\newblock In Andreas Krause, Emma Brunskill, Kyunghyun Cho, Barbara Engelhardt,
  Sivan Sabato, and Jonathan Scarlett, editors, {\em International Conference
  on Machine Learning, {ICML} 2023, 23-29 July 2023, Honolulu, Hawaii, {USA}},
  volume 202 of {\em Proceedings of Machine Learning Research}, pages
  24950--24962. {PMLR}, 2023.

\bibitem{mitrovic2023chatgpt}
Sandra Mitrovic, Davide Andreoletti, and Omran Ayoub.
\newblock Chatgpt or human? detect and explain. explaining decisions of machine
  learning model for detecting short chatgpt-generated text.
\newblock {\em CoRR}, abs/2301.13852, 2023.

\bibitem{GPTSniffer}
Phuong~T. Nguyen, Juri~Di Rocco, Claudio~Di Sipio, Riccardo Rubei, Davide~Di
  Ruscio, and Massimiliano~Di Penta.
\newblock Gptsniffer: {A} codebert-based classifier to detect source code
  written by chatgpt.
\newblock {\em J. Syst. Softw.}, 214:112059, 2024.

\bibitem{empirical_LLMs}
Changan Niu, Chuanyi Li, Vincent Ng, Dongxiao Chen, Jidong Ge, and Bin Luo.
\newblock An empirical comparison of pre-trained models of source code.
\newblock In {\em 45th {IEEE/ACM} International Conference on Software
  Engineering, {ICSE} 2023, Melbourne, Australia, May 14-20, 2023}, pages
  2136--2148. {IEEE}, 2023.

\bibitem{chatGPT_code_det_oedingen}
Marc Oedingen, Raphael~C. Engelhardt, Robin Denz, Maximilian Hammer, and
  Wolfgang Konen.
\newblock Chatgpt code detection: Techniques for uncovering the source of code.
\newblock {\em CoRR}, abs/2405.15512, 2024.

\bibitem{oman1989FirstStyl}
Paul~W. Oman and Curtis~R. Cook.
\newblock Programming style authorship analysis.
\newblock In Arthur~M. Riehl, editor, {\em Computer Trends in the 1990s -
  Proceedings of the 1989 {ACM} 17th Annual Computer Science Conference,
  Louisville, Kentucky, USA, February 21-23, 1989}, pages 320--326. {ACM},
  1989.

\bibitem{tiobe}
Marvin~Wener Paul~Jansen, Rob~Goud.
\newblock Tiobe - the software quality company.
\newblock \url{https://www.tiobe.com/tiobe-index/}.
\newblock Accessed: 2024-05-13.

\bibitem{radford2019language}
Alec Radford, Jeffrey Wu, Rewon Child, David Luan, Dario Amodei, Ilya
  Sutskever, et~al.
\newblock Language models are unsupervised multitask learners.
\newblock {\em OpenAI blog}, 1(8):9, 2019.

\bibitem{claude_detect}
Musfiqur Rahman, SayedHassan Khatoonabadi, Ahmad Abdellatif, and Emad Shihab.
\newblock Automatic detection of llm-generated code: A case study of claude 3
  haiku.
\newblock {\em arXiv preprint arXiv:2409.01382}, 2024.

\bibitem{ren2024copyright}
Jie Ren, Han Xu, Pengfei He, Yingqian Cui, Shenglai Zeng, Jiankun Zhang,
  Hongzhi Wen, Jiayuan Ding, Hui Liu, Yi~Chang, and Jiliang Tang.
\newblock Copyright protection in generative {AI:} {A} technical perspective.
\newblock {\em CoRR}, abs/2402.02333, 2024.

\bibitem{codellama}
Baptiste Rozi{\`{e}}re, Jonas Gehring, Fabian Gloeckle, Sten Sootla, Itai Gat,
  Xiaoqing~Ellen Tan, Yossi Adi, Jingyu Liu, Tal Remez, J{\'{e}}r{\'{e}}my
  Rapin, Artyom Kozhevnikov, Ivan Evtimov, Joanna Bitton, Manish Bhatt,
  Cristian Canton{-}Ferrer, Aaron Grattafiori, Wenhan Xiong, Alexandre
  D{\'{e}}fossez, Jade Copet, Faisal Azhar, Hugo Touvron, Louis Martin, Nicolas
  Usunier, Thomas Scialom, and Gabriel Synnaeve.
\newblock Code llama: Open foundation models for code.
\newblock {\em CoRR}, abs/2308.12950, 2023.

\bibitem{sandoval2023code_llm_security}
Gustavo Sandoval, Hammond Pearce, Teo Nys, Ramesh Karri, Siddharth Garg, and
  Brendan Dolan{-}Gavitt.
\newblock Lost at {C:} {A} user study on the security implications of large
  language model code assistants.
\newblock In Joseph~A. Calandrino and Carmela Troncoso, editors, {\em 32nd
  {USENIX} Security Symposium, {USENIX} Security 2023, Anaheim, CA, USA, August
  9-11, 2023}, pages 2205--2222. {USENIX} Association, 2023.

\bibitem{attentionIsAllYouNeed}
Ashish Vaswani, Noam Shazeer, Niki Parmar, Jakob Uszkoreit, Llion Jones,
  Aidan~N. Gomez, Lukasz Kaiser, and Illia Polosukhin.
\newblock Attention is all you need.
\newblock In Isabelle Guyon, Ulrike von Luxburg, Samy Bengio, Hanna~M. Wallach,
  Rob Fergus, S.~V.~N. Vishwanathan, and Roman Garnett, editors, {\em Advances
  in Neural Information Processing Systems 30: Annual Conference on Neural
  Information Processing Systems 2017, December 4-9, 2017, Long Beach, CA,
  {USA}}, pages 5998--6008, 2017.

\bibitem{wang2023codet5+}
Yue Wang, Hung Le, Akhilesh Gotmare, Nghi D.~Q. Bui, Junnan Li, and Steven
  C.~H. Hoi.
\newblock Codet5+: Open code large language models for code understanding and
  generation.
\newblock In Houda Bouamor, Juan Pino, and Kalika Bali, editors, {\em
  Proceedings of the 2023 Conference on Empirical Methods in Natural Language
  Processing, {EMNLP} 2023, Singapore, December 6-10, 2023}, pages 1069--1088.
  Association for Computational Linguistics, 2023.

\bibitem{codeDetectGPT_yang}
Xianjun Yang, Kexun Zhang, Haifeng Chen, Linda~R. Petzold, William~Yang Wang,
  and Wei Cheng.
\newblock Zero-shot detection of machine-generated codes.
\newblock {\em CoRR}, abs/2310.05103, 2023.

\bibitem{ziegler2021copilot_recitation}
Albert Ziegler.
\newblock {GitHub} {Copilot} research recitation --- parrot or crow? a first
  look at rote learning in {GitHub} {Copilot} suggestions.
\newblock
  \url{https://github.blog/ai-and-ml/github-copilot/github-copilot-research-recitation/},
  2021.
\newblock Accessed: 2024-09-25.

\end{thebibliography}


\clearpage
\end{document}